\documentclass[aip,prb,amsmath,amssymb,showpacs,manuscript]{revtex4}
\usepackage{graphicx}
\usepackage{bm}
\usepackage{}

\begin{document}

\title{Gap-modulated doping effects on indirect exchange
interaction between magnetic impurities in graphene}

\author{O. Roslyak and Godfrey Gumbs}

\affiliation{Department of Physics and Astronomy,
Hunter College, City University of New York,\\
695 Park Avenue, New York, NY 10065, USA}

\author{Danhong Huang}
\affiliation{Air Force Research Laboratory, Space Vehicles
Directorate,\\
Kirtland Air Force Base, NM 87117, USA}

\date{\today}

\begin{abstract}
A dilute distribution of magnetic impurities  is assumed
to be present in doped graphene. We calculate the interaction
energy  between two magnetic impurities which are coupled
via the indirect-exchange or Ruderman-Kittel-Kasuva-Yosida (RKKY)
interaction by the doped conduction electrons. The current model
is a half-filled $AB$-lattice structure. Our calculations are based
on the  retarded lattice Green's function formalism in
momentum-energy space which is employed in linear response
theory to determine the magnetic susceptibility in coordinate space.
Analytic results are obtained for gapped graphene when the
magnetic impurities are placed on the $A$ and $B$ sublattice
sites of the structure. This interaction, which is important
in determining spin ordering, has been found to be
significantly different for $AA$ and $BB$ exchange energies
in doped graphene due to the existence of an energy gap,
and is attributed to a consequence of the local fields
not being equal on the $A$ and $B$ sublattices.
For doped graphene, the oscillations of all three RKKY
interactions from ferromagnetic to antiferromagnetic with
increasing Fermi energy is significantly modified by the
energy gap both in magnitude and  phase. Additionally, the $AB$
exchange energy may be modified by the presence of a gap for
undoped graphene but not for doped graphene due to the dominance
of doped conduction electrons.
\end{abstract}

                             % Classification Scheme.

\maketitle

\section{Introduction}
\label{SEC:1}

Recently, there has been some attention given to the effects arising from
the  Ruderman-Kittel-Kasuva-Yosida (RKKY)\,\cite{RKKY1,RKKY2,RKKY3} or indirect-exchange  interaction between spins via the host conduction
electrons of monolayer free standing graphene,  a two-dimensional
honeycomb network of carbon atoms\,\cite{RKKYG1,RKKYG2,RKKYG3,RKKYG4,RKKYG5,RKKYG6,RKKYG7,RKKYG8}.
Such local moments may arise  near extended defects.
Interest in the RKKY  interaction is due partially  to the fact that it
determines spin ordering. The indirect exchange interaction between local
magnetic moments is generally determined by electron excitations near
the Fermi level. However, although the RKKY interaction has been considered
in intrinsic graphene by several authors, little attention has been
given to its role when a gap $2E_g$ is opened at the two inequivalent
$K$ and $K^\prime$ points in momentum space between the valence  and
conduction bands\,\cite{pavlo,7,8,kibis,OR1,OR2}. This effective
band gap may be generated by spin-orbit interaction\,\cite{RKKYG7}.
It may also arise  when monolayer graphene is placed on a substrate such as
ceramic silicon carbide\,\cite{7} or graphite\,\cite{8} or
dynamically when it is irradiated by circularly polarized
light\,\cite{kibis,OR1,OR2}. Depending on the nature of the
substrate  on which graphene is placed or the intensity and
amplitude of the light, the gap may be a few meV or as large as an eV. In
general, the energy gap is attributed to a breakdown in symmetry between the
sublattices caused by external perturbing fields from the substrate or photons
coupled to the $A$ and $B$ atoms. Furthermore, there still exists a small
band gap $2E_{\rm so}\sim 10^{-3}$ meV  due to spin-orbit
coupling\,\cite{9,9b} even in the absence of a substrate or an
external laser  field. Despite the formation of an energy band
gap, corresponding to a metal-insulator transition,  for the
half-filled bipartite lattice, the interaction between atoms
on the same lattice has still been suggested to be ferromagnetic
and antiferromagnetic between atoms on different
sublattices in \emph{undoped} graphene\,\cite{RKKYG6}, even though
the energy dispersion is no longer linear and isotropic
as it is for gapless Dirac electrons. For undoped graphene,
we show that the existence of an energy gap
can significantly modify the magnitude of the magnetic interaction
between impurities on the sublattices. More interestingly, for
doped graphene, we demonstrate that the energy gap
can drastically affect the nature of such a magnetic interaction.
\medskip

In Ref.\,[\onlinecite{RKKYG7}], with a small gap at the Fermi level
due to spin-orbit interaction, the graphene layer was assumed to be undoped.
The present paper  explores quantitatively the role played by massive
relativistic Dirac particles in both doped and undoped graphene on the RKKY
indirect-exchange energy. In other words, how the magnetic effects
depend on the chirality of the electron states for doped graphene
with an energy gap. Analytic results are obtained for gapped graphene
when both impurities are located on sites belonging to the $A$
sublattice, both on the  $B$ sublattice and when one impurity
is on an $A$ while the other is on a $B$ sublattice site. We
demonstrate numerically the large asymmetry between the $AA$ and $BB$-exchange-interaction energy  for gapped graphene. Our closed-form
analytic results make it convenient for our predictions to be
compared with experimental data once they become available.
\medskip

The remainder of this paper is organized as follows. In
Sec.\,\ref{SEC:2}, we obtain the  retarded Green's function matrix
elements on the $A$ and $B$ sublattices for gapped graphene.
In Sec.\,\ref{SEC:3}, we use linear response theory to calculate
the magnetic susceptibility by making use of our derived results
for the Green's functions on the bipartite sublattices. The
interaction energy of two spins at lattice sites $r_\mu$ and
$r_\nu^\prime$, where $\mu,\,\nu=A,\,B$, is then deduced for both
undoped and doped gapped graphene. We conclude our paper in
Sec.\,\ref{SEC:4} with some remarks.

\section{Green-Function Formalism for Graphene}
\label{SEC:2}

The Hamiltonian for  monolayer gaped graphene in the
absence of impurities has the form:

\begin{equation}
\label{EQ:7}
{\cal H}(\mathbf{k})=
\left[{
\begin{matrix}
E_g & -\gamma_0 h_0^\star(\mathbf{k})\\
-\gamma_0 h_0(\mathbf{k}) & -E_g
\end{matrix}
}\right]\ ,
\end{equation}
where

\begin{equation}
h_0(\mathbf{k})= \exp\left({i k_y a}\right)
+\exp\left({-\frac{i \sqrt{3} a k_x}{2} -\frac{i a k_y}{2}}\right)
+\exp\left({\frac{i \sqrt{3} a k_x}{2} -\frac{i a k_y}{2}}\right)
\end{equation}
with the carbon-carbon distance $a= 1.14$\,\AA, $\gamma_0 = - 3.0\,\texttt{eV}$ and $\mathbf{k}=(k_x,k_y)$.
Here, $2E_g$ is the energy gap generated by some means, possibly by  a substrate or
circularly polarized light, as discussed in the Introduction.
The retarded Green's function   matrix in wave vector-energy ($\mathbf{k},E$) space
is given by

\begin{eqnarray}
\label{EQ:8}
G\left({\mathbf{k},E}\right)&=&
\left[{E-{\cal H}(\mathbf{k})+ i 0^+}\right]^{-1}
\nonumber\\
&=&\frac{1}{E^2-E_g^2 - \gamma_0^2\,h_0(\mathbf{k})h_0^\star(\mathbf{k})}
\left[{ \begin{matrix}
E_g+E & \gamma_0 h_0^\star(\mathbf{k})\\
 \gamma_0 h_0(\mathbf{k}) & - E_g + E
\end{matrix}
}\right]\equiv \left[{ \begin{matrix}
G_{AA} & G^\star_{AB}\\
G_{AB} & G_{BB}
\end{matrix}
}\right]
\end{eqnarray}
with

\begin{eqnarray}
G_{AA} \left({\mathbf{k},E}\right) &=& \frac{1}{E^2
-E^2_{\mathbf{k}}}\,(E_g + E)\ ,
\nonumber\\
G_{BB} \left({\mathbf{k},E}\right) &=& \frac{1}{E^2
-E^2_{\mathbf{k}}}\,(-E_g + E)\ ,
\nonumber\\
G_{AB} \left({\mathbf{k},E}\right)  &=& G^\star_{BA} \left({\mathbf{k},E}\right)=\frac{1}{E^2-E^2_{\mathbf{k}}}\,
\gamma_0h_0(\mathbf{k})\ ,
\nonumber\\
\end{eqnarray}
where we introduced the energy dispersion equation $E^2_{\mathbf{k}} = \gamma_0^2\,h_0(\mathbf{k})\,h_0^\star(\mathbf{k})+E^2_g$.
\medskip

We may now transform the Green's function to the real space representation
with use of

\begin{eqnarray}
G(\mathbf{r},\mathbf{r}^\prime,E) &=&
\int \limits_{\texttt{1$^{\rm st}$\,BZ}} d^2 \mathbf{k} \
G(\mathbf{k},E) \exp\left[{
i\mathbf{k}\cdot\left({\mathbf{r}-\mathbf{r}^\prime}\right)
}\right]
\nonumber\\
&=& \frac{S}{4 \pi^2}
\sum \limits_{i=1}^{6}
\int \limits dk_x dk_y\,G(\mathbf{K}_i + \mathbf{k},E)
\exp \left[{
i\left({\mathbf{K}_i + \mathbf{k} }\right)\cdot\left({\mathbf{r}-\mathbf{r}^\prime}\right)
}\right]\ ,
\end{eqnarray}
where the summation is carried out  over the six corners of the
Brillouin zone (BZ), so that we may make the approximation
$E^2_{\mathbf{K}_i + \mathbf{k} } =
\left({\gamma k}\right)^2 + E^2_g$ with $\gamma = 3 a \gamma_0 /2$.
In that approximation, we also have $\gamma_0h_0 = \gamma k$.
Additionally, in our notation, the area of the unit cell
is $S= 3 \sqrt{3} a^2/2$.
\medskip

Using cylindrical coordinates, the matrix elements
of the retarded Green's function may be  explicitly written as

\begin{eqnarray}
G_{AA}(\textbf{r}_A,\textbf{r}^\prime_A;E)&=&
{\frac{2S}{(2\pi)^2}}\,\cos[\textbf{K}_1\cdot(\textbf{r}_A-\textbf{r}^\prime_A)]
\int\limits_0^{2\pi}d\theta\int\limits_0^{k_c}dk\ k
{G_{AA}(k,E)}\,e^{i\textbf{k}\cdot(\textbf{r}_A-\textbf{r}^\prime _A)}\ ,
\nonumber\\
G_{BB}(\textbf{r}_B,\textbf{r}^\prime_B;E)&=&{\frac{2S}{(2\pi)^2}}\,\cos[\textbf{K}_1\cdot(\textbf{r}_B
-\textbf{r}^\prime_B)]
\int\limits_0^{2\pi}d\theta\int\limits_0^{k_c}dk\ k
{G_{BB}(k,E)}\,e^{i\textbf{k}\cdot(\textbf{r}_B-\textbf{r}^\prime_B)}\ ,
\nonumber\\
G_{AB}(\textbf{r}_A,\textbf{r}^\prime_B;E)&=&G^\star_{BA}(\textbf{r}^\prime_B,\textbf{r}_A;E)
\nonumber\\
&=&{\frac{S}{(2\pi)^2}}\{
e^{i\textbf{K}_1(\textbf{r}_A-\textbf{r}^\prime_B)}
\int\limits_0^{2\pi}d\theta\int\limits_0^{k_c}dk\ k(-i\sin\theta-\cos\theta)
{G_{AB}(k,E)}\,e^{i\textbf{k}\cdot(\textbf{r}_A-\textbf{r}^\prime_B)}
\nonumber\\
&+&e^{-i\textbf{K}_1(\textbf{r}_A-\textbf{r}^\prime_B)}
\int\limits_0^{2\pi}d\theta
\int\limits_0^{k_c}dk\  k(-i\sin\theta+\cos\theta)
{G_{AB}(k,E)}\,e^{i\textbf{k}\cdot(\textbf{r}_A-\textbf{r}^\prime_B)}\}\ ,
\end{eqnarray}
where $k_c$ is a cut-off wave vector whose existence is due to the fact that
the energy band structure used in the calculation is only valid for a limited range
of the first BZ near the $K$ points. However, it is also allowed in such an
approximation to take the limit  $k_c \to \infty$ from a mathematical point of view.
In the above expression, we have $\textbf{K}_1=(4\pi/3\sqrt{3}a,0)$.
\medskip

We now make use of the following identity involving the series of the first kind of Bessel function $J_n(x)$

\begin{equation}
\label{EQ:9}
e^{i\textbf{k}\cdot(\textbf{r}_\mu-\textbf{r}^\prime_\nu)}
=J_0(k|\textbf{r}_\mu-\textbf{r}^\prime_\nu|)
+2\,{\sum\limits_{n=1}^\infty}\,i^nJ_n(k|\textbf{r}_\mu-\textbf{r}^\prime_\nu|)
\cos(n\varphi_{r_\mu,r^\prime_\nu})\ ,
\end{equation}
where $(\mu,\nu)\equiv(A,B)$ and $\varphi_{r_\mu,r^\prime_\nu}$ is the
angle between $\textbf{k}$ and $\textbf{r}_\mu-\textbf{r}^\prime_\nu$.
We may calculate the above integrals and simplify the
real-space Green's function matrix elements as

\begin{eqnarray}
G_{AA}(\textbf{r}_A,\textbf{r}_A^\prime;E)&=&\cos[\textbf{K}_1\cdot(\textbf{r}_A
-\textbf{r}^\prime_A)]\,F_1(|\textbf{r}_A-\textbf{r}^\prime_A|,E)\ ,
\nonumber\\
G_{BB}(\textbf{r}_B,\textbf{r}_B^\prime;E)&=&\cos[\textbf{K}_1\cdot(\textbf{r}_B
-\textbf{r}^\prime_B)]\,F_2(|\textbf{r}_B-\textbf{r}^\prime_B|,E)\ ,
\nonumber\\
G_{AB}(\textbf{r}_A,\textbf{r}_B^\prime;E)&=&G^\star_{BA}(\textbf{r}_B^\prime,\textbf{r}_A;E)=\sin[\textbf{K}_1\cdot(\textbf{r}_A
-\textbf{r}^\prime_B)+\alpha_{r_A,r^\prime_B}]\,F_3(|\textbf{r}_A-\textbf{r}^\prime_B|,E)\ .
\end{eqnarray}
In this notation,
$\cos(\varphi_{r_\mu,r^\prime_\nu})=\cos(\theta-\alpha_{r_\mu,r^\prime_\nu})$,
where $\theta$ is the angle which $\textbf{k}$ makes with the $x$-axis, and
$\alpha_{r_\mu,r^\prime_\nu}$ is the angle between
$\textbf{r}_\mu-\textbf{r}^\prime_\nu$ and the $x$-axis. Three position-dependent complex functions
$F_j(|\textbf{r}_\mu-\textbf{r}^\prime_\nu|,E)$ for $j=1,\,2,\,3$ are defined as follows:

\begin{eqnarray}
\label{EQ:12}
F_1(|\textbf{r}_\mu-\textbf{r}^\prime_\nu|,E)&=&\frac{S}{\pi}
{\int\limits_0^{\infty}}dk\  kJ_0(k|\textbf{r}_\mu-\textbf{r}^\prime_\nu|)\,
\frac{E_g + E}{E^2-(\gamma k)^2 - E^2_g}
\nonumber\\
&=&-\frac{S}{\pi \gamma^2} \left[{
(E+E_g) K_0 \left({\frac{i \sqrt{E^2-E_g^2}}{\gamma}|
\textbf{r}_\mu-\textbf{r}^\prime_\nu|}\right)
}\right] \ ,
\nonumber\\
F_2(|\textbf{r}_\mu-\textbf{r}^\prime_\nu|,E)&=&\frac{S}{\pi}
{\int\limits_0^{\infty}}dk\  kJ_0(k|\textbf{r}_\mu-\textbf{r}^\prime_\nu|)\,
\frac{-E_g + E}{E^2-(\gamma k)^2 - E^2_g}
\nonumber\\
&=&-\frac{S}{\pi \gamma^2} \left[{ (E-E_g) K_0 \left({\frac{i \sqrt{E^2-E_g^2}}{\gamma}|\textbf{r}_\mu-\textbf{r}^\prime_\nu|}\right)
}\right]\ ,
\nonumber\\
F_3(|\textbf{r}_\mu-\textbf{r}^\prime_\nu|,E)&=&\frac{S}{\pi}
{\int\limits_0^{\infty}}dk\  kJ_1(k|\textbf{r}_\mu-\textbf{r}^\prime_\nu|)\,
\frac{\gamma k}{E^2-(\gamma k)^2 - E^2_g}
\nonumber\\
& = &-\frac{S}{\pi \gamma^2}\left[{ i \sqrt{E^2-E_g^2} K_1 \left({\frac{i \sqrt{E^2-E_g^2}}{\gamma}|\textbf{r}_\mu-\textbf{r}^\prime_\nu|}\right)
}\right] \ ,
\end{eqnarray}
where $K_m(x)$ for integer $m$ is the modified Bessel function. In addition,
we have used the formula for the $m^{\texttt{th}}$ order Hankel
transform

\begin{equation}
\int\limits_0^\infty \ xdx\
\frac{x^m}{x^2 +\alpha^2}\,J_m (xE)   = \alpha^m K_m (\alpha E)\ ,
\end{equation}
where $\alpha$ is an arbitrary constant.

\section{Indirect-Exchange Interaction Energy}
\label{SEC:3}

Let us now assume that we have two magnetic impurities with spins
$\mathbf{S}_1$ and $\mathbf{S}_2$) located at
$\mathbf{r}_\mu$ and $\mathbf{r^\prime}_\nu$, respectively. According
to  linear response theory, the energy needed to exchange
(mediated by the Dirac electrons) their positions may be written in the matrix form

\begin{equation}
{\cal V}_{\mu\nu}(|\textbf{r}_\mu-\textbf{r}^\prime_\nu|) = {\cal J}_{\mu\nu}(|
\textbf{r}_\mu-\textbf{r}^\prime_\nu|)\,\mathbf{S}_1 \cdot \mathbf{S}_2\ ,
\end{equation}
where the exchange-integral matrix is proportional to the matrix of spin-independent susceptibility

\begin{equation}
{\cal J}_{\mu\nu} (|\textbf{r}_\mu-\textbf{r}^\prime_\nu|) = \frac{4\lambda^2}{\hbar^2}\,\chi_{\mu\nu}(|\textbf{r}_\mu-\textbf{r}^\prime_\nu|)\ .
\end{equation}
Here, $\lambda$ is the contact interaction between the magnetic impurities and the
Dirac electrons. The magnetic susceptibility may be expressed in terms of
the Green's functions in the standard RKKY form\,\cite{RKKY1,RKKY2,RKKY3}.

\subsection{Undoped Graphene with a Gap}

For \emph{undoped} graphene, we take the Fermi energy $E_F=0$.
For this case, we obtain the matrix of spin-independent susceptibility

\begin{equation}
\chi_{\mu\nu}(|\textbf{r}_\mu-\textbf{r}^\prime_\nu|) =
-\frac{2}{\pi} \int\limits_{-\infty}^{-E_g} dE\  \texttt{Im}
\left[{G(\textbf{r}_\mu,\textbf{r}_\nu^\prime;E)\,
G(\textbf{r}_\nu^\prime,\textbf{r}_\mu;E) }\right] \ .
\end{equation}
Making use of the expressions for the Green's functions from the preceding
section, it is a simple matter  to obtain the following matrix elements:

\begin{eqnarray}
\chi_{AA}(|\textbf{r}_A-\textbf{r}^\prime_A|) &=& -\frac{1}{\pi}
\left({1+\cos[2\textbf{K}_1\cdot(\textbf{r}_A-\textbf{r}^\prime_A)]}\right)\,
\texttt{Im} \int \limits_{-\infty}^{-E_g} dE
\left[{F_1(|\textbf{r}_A-\textbf{r}^\prime_A|,E)}\right]^2\ ,
\nonumber\\
\chi_{BB}(|\textbf{r}_B-\textbf{r}^\prime_B|) &=& -\frac{1}{\pi}
\left({1+\cos[2\textbf{K}_1\cdot(\textbf{r}_B-\textbf{r}^\prime_B)]}\right)\,
\texttt{Im} \int \limits_{-\infty}^{-E_g} dE
\left[{F_2(|\textbf{r}_B-\textbf{r}^\prime_B|,E)}\right]^2\ ,
\nonumber\\
\chi_{AB}(|\textbf{r}_A-\textbf{r}^\prime_B|) &=& -\frac{1}{\pi}
\left({\cos[2\textbf{K}_1\cdot(\textbf{r}_A-\textbf{r}^\prime_B)
+ 2 \alpha_{r_A,r^\prime_B}]-1}\right)
\nonumber\\
&\times &
\texttt{Im} \int \limits_{-\infty}^{-E_g} dE
\left[{F_3(|\textbf{r}_A-\textbf{r}^\prime_B|,E)}\right]^2\ .
\end{eqnarray}
We  first consider the case when both  impurities are located on
$A$ atomic sites. That is, we need to calculate

\begin{eqnarray}
&& {\cal P}_1(E_g, R_{AA}) \equiv \texttt{Im} \int \limits_{-\infty}^{-E_g} dE
\left[{F_1(|\textbf{r}_A-\textbf{r}^\prime_A|,E)}\right]^2
\nonumber\\
&=&
\frac{S^2}{2 \gamma^4}\int \limits_{-\infty}^{-E_g} dE \
\left({E+E_g}\right)^2
J_0 \left({\frac{- \sqrt{E^2-E_g^2}}{\gamma}\,|\textbf{r}_A-\textbf{r}^\prime_A|}\right)
N_0 \left({\frac{- \sqrt{E^2-E_g^2}}{\gamma}\,|\textbf{r}_A-\textbf{r}^\prime_A|}\right)
\nonumber\\
&=&  \frac{2}{3 \gamma_0}\int \limits_{-\infty}^{-E_g} dE \
\left({E+E_g}\right)^2
J_0 \left({- \frac{2}{3} \sqrt{E^2-E_g^2}\,R_{AA}}\right)
N_0 \left({- \frac{2}{3} \sqrt{E^2-E_g^2}\,R_{AA}}\right)\ ,
\end{eqnarray}
where $N_0(x)$ is  the Neumann
function. For convenience,  in the last expression, we measure
energy in   units of $\gamma_0$ and distance in  units of $R_{AA}=|\textbf{r}_A-\textbf{r}^\prime_A|/a$. In the gapless case,
$E_g=0$, we are able to  calculate the integral analytically, obtaining

\begin{eqnarray}
\label{EQ:F1}
&& {\cal P}^0_1(R_{AA})\equiv{\cal P}_1(0, R_{AA}) = \texttt{Im} \int \limits_{-\infty}^{0} dE \
\left[{F_1(|\textbf{r}_A-\textbf{r}^\prime_A|,E)}\right]^2
\nonumber\\
&=&\frac{9}{4 \gamma_0 R^3_{AA}}\int \limits_{0}^{\infty} dx\,
x^2\,
J_0 \left({x}\right)
N_0 \left({x}\right)=\frac{9}{64 \gamma_0 R^3_{AA}} \ .
\end{eqnarray}
The last integral may be evaluated through

\begin{eqnarray}
&& \lim_{x_0 \rightarrow 0} \int \limits_{0}^{\infty} dx\,
x^2 \exp(- x_0 x)\,
J_0 \left({x}\right)
N_0 \left({x}\right)
\nonumber\\
&=&
\lim_{x_0 \to 0} \frac{4 \left({4+3 x^2_0}\right)
{\cal E}(-x_0^2/4)-2 \left({8+6 x_0^2 +x_0^4}\right) {\cal K}(-x_0^2/4)}
{\pi x_0^2 \left({4+x_0^2}\right)^2} =\frac{1}{16} \ ,
\end{eqnarray}
where ${\cal K}(x)$, ${\cal E}(x)$ are the  complete elliptic integrals of the
first and second kind, respectively.
\medskip

In a similar fashion, we have for the $AB$ exchange interaction

\begin{eqnarray}
{\cal P}_3(E_g, R_{AB}) &=& \texttt{Im} \int \limits_{-\infty}^{-E_g} dE
\left[{F_3(|\textbf{r}_A-\textbf{r}^\prime_B|,E)}\right]^2=-\frac{2}{3 \gamma_0} \int \limits_{-\infty}^{-E_g} dE\,
\left(\sqrt{E^2-E_g^2}\right)^2
\nonumber\\
&\times&
J_1 \left({- \frac{2}{3} \sqrt{E^2-E_g^2}R_{AB}}\right)
N_1 \left({- \frac{2}{3} \sqrt{E^2-E_g^2}R_{AB}}\right) \ ,
\end{eqnarray}
where $N_1(x)$ is also the Neumann function.
For $E_g =0$, the integral may be evaluated analytically, yielding

\begin{eqnarray}
&& \lim_{x_0 \to 0} \int \limits_{0}^{\infty} dx\
x^2 \exp(- x_0 x)\,
J_1 \left({x}\right)
N_1 \left({x}\right)
\nonumber\\
&=& \lim_{x_0 \rightarrow 0} \frac{-4
\left({-4+ x^2_0}\right) {\cal E}(-x_0^2/4)-4 \left({4+ x_0^2}\right) {\cal K}(-x_0^2/4)}
{\pi x_0^2 \left({4+x_0^2}\right)^2} =-\frac{3}{16}\ ,
\end{eqnarray}
so that we have

\begin{equation}
\label{EQ:F3}
{\cal P}^0_3(R_{AB})\equiv{\cal P}_3(0, R_{AB})=\texttt{Im} \int \limits_{-\infty}^{0} dE
\left[{F_3(|\textbf{r}_A-\textbf{r}^\prime_B|,E)}\right]^2
=-\frac{27}{64 \gamma_0 R^3_{AB}} \ .
\end{equation}
\medskip

The  effects of doping on the indirect-exchange interaction for
a graphene layer without a gap have been studied in detail.
In this case,  a sign change of the indirect-exchange interaction
was discovered when  the doping concentration was varied.
 There is a switching from inverse cubic  to inverse square
 power law for the spatial dependence.\,\cite{RKKYG5}
However, the doping effect for a graphene layer with a gap
remains largely unexplored. To demonstrate the doping effect
for gaped graphene, we write ${\cal P}_j(E_g, R_{\mu\nu})={\cal P}^0_j(R_{\mu\nu})+\Delta {\cal P}_j(E_g, R_{\mu\nu})$ with $\mu,\,\nu=A,\,B$ and $j=1,\,2,\,3$.
\medskip

In Fig.\,\ref{FIG:1}, we present our calculated results for
${\cal P}_j(E_g, R_{\mu\nu})$ (left panel) and
$\Delta {\cal P}_j(E_g, R_{\mu\nu})$ (right panel) in an
undoped graphene layer as functions $E_g$ with $R_{AA}=R_{BB}=R_{AB}=0.5$.
From the left panel of this figure, we find that the positive
exchange interactions ${\cal P}_1$ and ${\cal P}_2$ between
two intra-sublattice impurities (intra-SIs) decrease with
$E_g$ while the negative exchange interaction ${\cal P}_3$
between two inter-sublattice impurities (inter-SIs) increases
with $E_g$ at the same time. We further observe that the degenerate
 ${\cal P}_1$ and ${\cal P}_2$ at $E_g=0$, due to the presence
  of  symmetry with respect to two sublattices, is now
  lifted  for $E_g>0$. Compared with the results when $E_g=0$.
We see from the right panel of this figure that the gap-induced
changes $\Delta {\cal P}_1$ and $\Delta {\cal P}_2$ become
more and more negative with increasing $E_g$. Additionally,
the decrease of $\Delta {\cal P}_1$ is more rapid than
for $\Delta {\cal P}_2$. This behavior reflects a modulation
in the exchange interaction by an energy gap in graphene. Also,
it suggests a possible gap-induced ferromagnetic-to-antiferromagnetic
phase transition due to the intra-SI exchange interaction.
In contrast, a distinctly opposite behavior is found in the
inter-SI exchange interaction $\Delta {\cal P}_3$.
\medskip

\begin{figure}[ht]
\centering
\includegraphics[width=\textwidth]{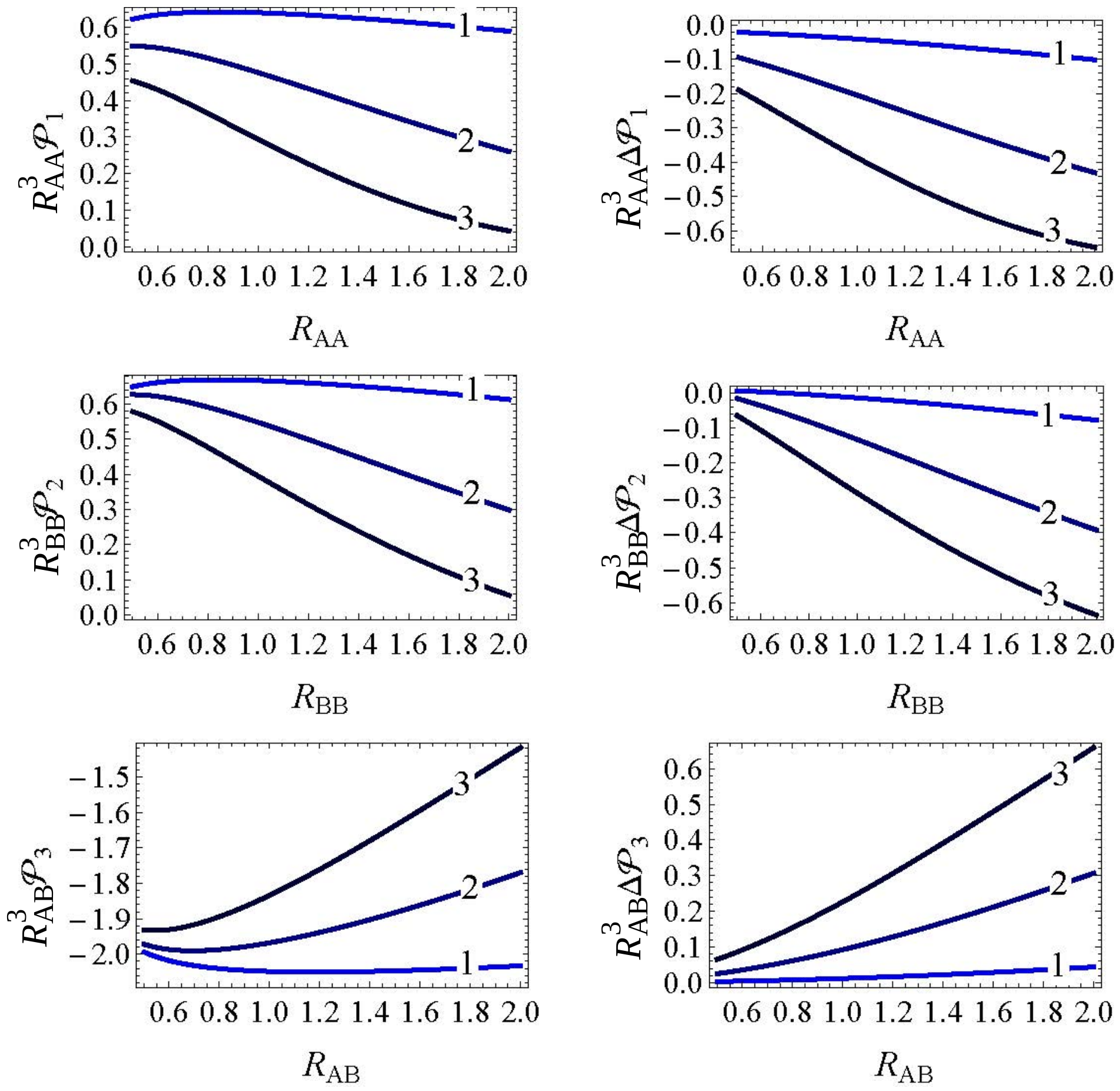}
\caption{\label{FIG:2} (Color online) Exchange interaction energies $R_{AA}^3{\cal P}_1$ (top panel), $R_{BB}^3{\cal P}_2$ (middle panel) and $R_{AB}^3{\cal P}_3$ (bottom panel) on the left column,
as well as the changes of them $R_{AA}^3\Delta{\cal P}_1$ (top panel), $R_{BB}^3\Delta{\cal P}_2$ (middle panel) and $R_{AB}^3\Delta{\cal P}_3$ (bottom panel) on the right column,
in undoped graphene as functions of the distance between two magnetic impurities. Curves $1,\,2,\,3$ correspond to gap energy
parameter $E_g = 0.1,\ 0.3,\ 0.5$, respectively.
Here, we measure  energy in  units of $\gamma_0$ and
distance in units of $R_{\mu\nu}=|\textbf{r}_\mu-\textbf{r}^\prime_\nu|/a$ for $\mu,\,\nu=A,\,B$.}
\end{figure}

We display in Fig.\,\ref{FIG:2} the calculated results for
$R^3_{\mu\nu}{\cal P}_j(E_g, R_{\mu\nu})$ (left column) and
$R^3_{\mu\nu}\Delta {\cal P}_j(E_g, R_{\mu\nu})$ (right column)
as functions of the distance $R_{\mu\nu}$ between two magnetic
impurities for various values of $E_g$ in undoped graphene. From the
left column of this figure, we observe  clear deviations of
the exchange  interactions ${\cal P}_1$, ${\cal P}_2$ and
${\cal P}_3$ from $1/R^3_{\mu\nu}$ dependence (at $E_g=0$)
with increasing $E_g$ (from curve-1 to curve-3). These results
also indicate a possible gap-induced phase transition which may
 be driven by either the intra-SI or the inter-SI exchange
 interaction at a relatively large distance between two
 magnetic impurities in a graphene layer. Additionally, we
 find from the right column of this figure that changes in
 $\Delta {\cal P}_1$ and $\Delta {\cal P}_2$ in the intra-SI
 exchange interactions drop more rapidly as a function of
 $R_{\mu\mu}$ when $E_g$ is increased. Again, a completely
 opposite behavior is observed in this case for $\Delta {\cal P}_3$
which is associated with the inter-SI exchange interaction.

\subsection{Doped Graphene with a Gap}

We now turn to the case of doped graphene
which has a finite Fermi energy $E_F>0$. It is convenient to
introduce a dimensionless f the Fermi energy as
$X_F \equiv \frac{2}{3}E_F R_{AA}/\gamma_0$. The Fermi energy
makes the integration limits in Eqs.~\eqref{EQ:F1} and
\eqref{EQ:F3} raised to include the part
$\int \limits_{E_g}^{E_g+E_F} dE$. In the
case when $E_g = 0 $ those integrals assume the analytical forms:

\begin{gather}
{\cal P}^0_{1}\left({R_{AA}}\right) = \frac{9}{4 \gamma_0 R^3_{AA}}
\left[{
\frac{1}{16} +
\frac{{\cal G}_{2,4}^{2,1}\left(X_F^2\bigg|{
\begin{array}{c}
 1,2 \\
 \frac{3}{2},\frac{3}{2},0,\frac{3}{2} \\
\end{array}}
\right)}{2 \sqrt{\pi }}
}\right]\ ,\\
{\cal P}^0_{3}\left({R_{AB}}\right) = \frac{9}{4 \gamma_0 R^3_{AB}}
\left[{
-\frac{3}{16} -
\frac{{\cal G}_{2,4}^{2,1}\left(X_F^2\bigg|{
\begin{array}{c}
 1,2 \\
 \frac{3}{2},\frac{5}{2},0,\frac{1}{2} \\
\end{array}}
\right)}{2 \sqrt{\pi }}
}\right]\ .
\end{gather}
Here, ${\cal G}^{m,n}_{p,q}$ is the Meijer's $G$-function.
It has oscillating behavior\,\footnote{Bessel functions are
special cases of the general Meijer functions} thus switching
the nature of ${\cal J}_{AA}$ and ${\cal J}_{AB}$
between ferromagnetic and antiferromagnetic. In the long distance
limit $R_{AA}\approx R_{AB}\approx R_{BB}$, we can use the mean
field approximation as in the work of Sherafati and  Satpathy\,\cite{RKKYG3}.
\medskip

\begin{figure}[ht]
\centering
\includegraphics[width=\textwidth]{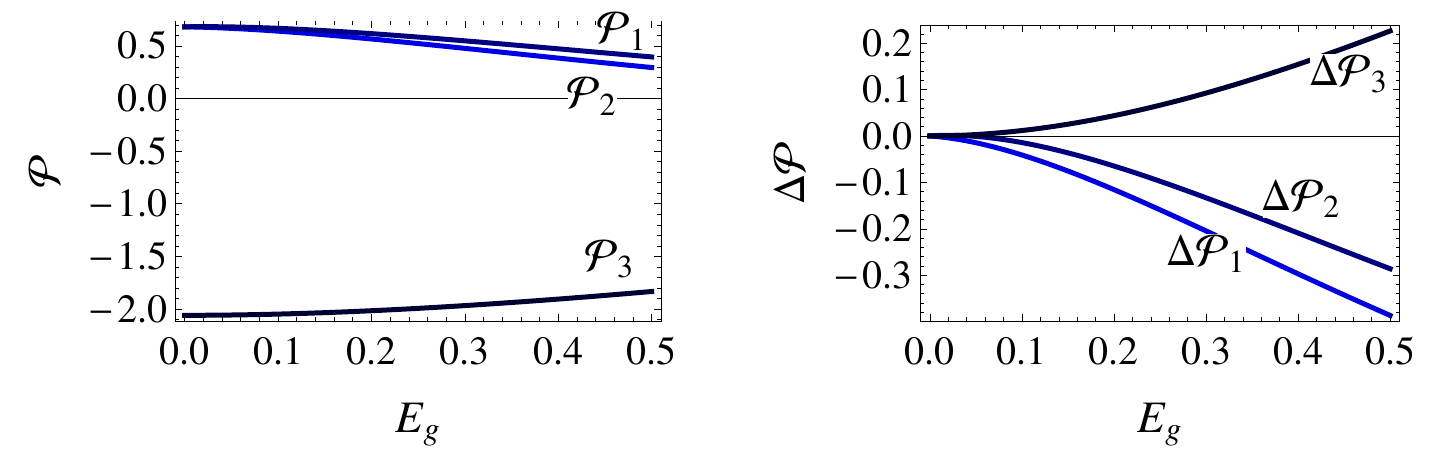}
\caption{\label{FIG:1} (Color online) Exchange interaction energies ${\cal P}_1$, ${\cal P}_2$ and ${\cal P}_3$ (left panel),
as well as the changes of them $\Delta{\cal P}_1$, $\Delta{\cal P}_2$ and $\Delta{\cal P}_3$ (right panel),
in undoped graphene as functions of the half-gap $E_g$ for fixed $R_{AA}=R_{BB}=R_{AB} =0.5$.
Here, we measure  energy in  units of $\gamma_0$ and
distance in units of $R_{\mu\nu}=|\textbf{r}_\mu-\textbf{r}^\prime_\nu|/a$ for $\mu,\,\nu=A,\,B$.}
\end{figure}

In Fig.\,\ref{FIG:3}, we present our calculated results for
${\cal P}_j(E_g, R_{\mu\nu})$ (left panel) and
$\Delta {\cal P}_j(E_g, R_{\mu\nu})$ (right panel) as well
as functions of $E_g$ with $R_{AA}=R_{BB}=R_{AB}=1$ and $E_F-E_g=1$
in doped graphene. The left panel of this figure shows
that the intra-SI exchange interaction ${\cal P}_1$ (${\cal P}_2$)
decreases (increases) with $E_g$ while the inter-SI exchange
interaction ${\cal P}_3$ is almost independent of  $E_g$ at
the same time. The anomalous increasing ${\cal P}_2$ with $E_g$ is in strong contrast with the results shown in the left panel of Fig.\,\ref{FIG:1}
and is indicative of  the band-filling effect on the gap-modulation
of the indirect exchange interaction between two magnetic impurities.
Moreover, from the right panel of this figure, we observe that
the gap-induced changes $\Delta {\cal P}_1$ and $\Delta {\cal P}_2$
acquire a linear dependence on $E_g$ in the presence of doping,
which is also in contrast with the results presented
in the right panel of Fig.\,\ref{FIG:1} where a nonlinear dependence
on $E_g$ is obtained.
\medskip

\begin{figure}[ht]
\centering
\includegraphics[width=\textwidth]{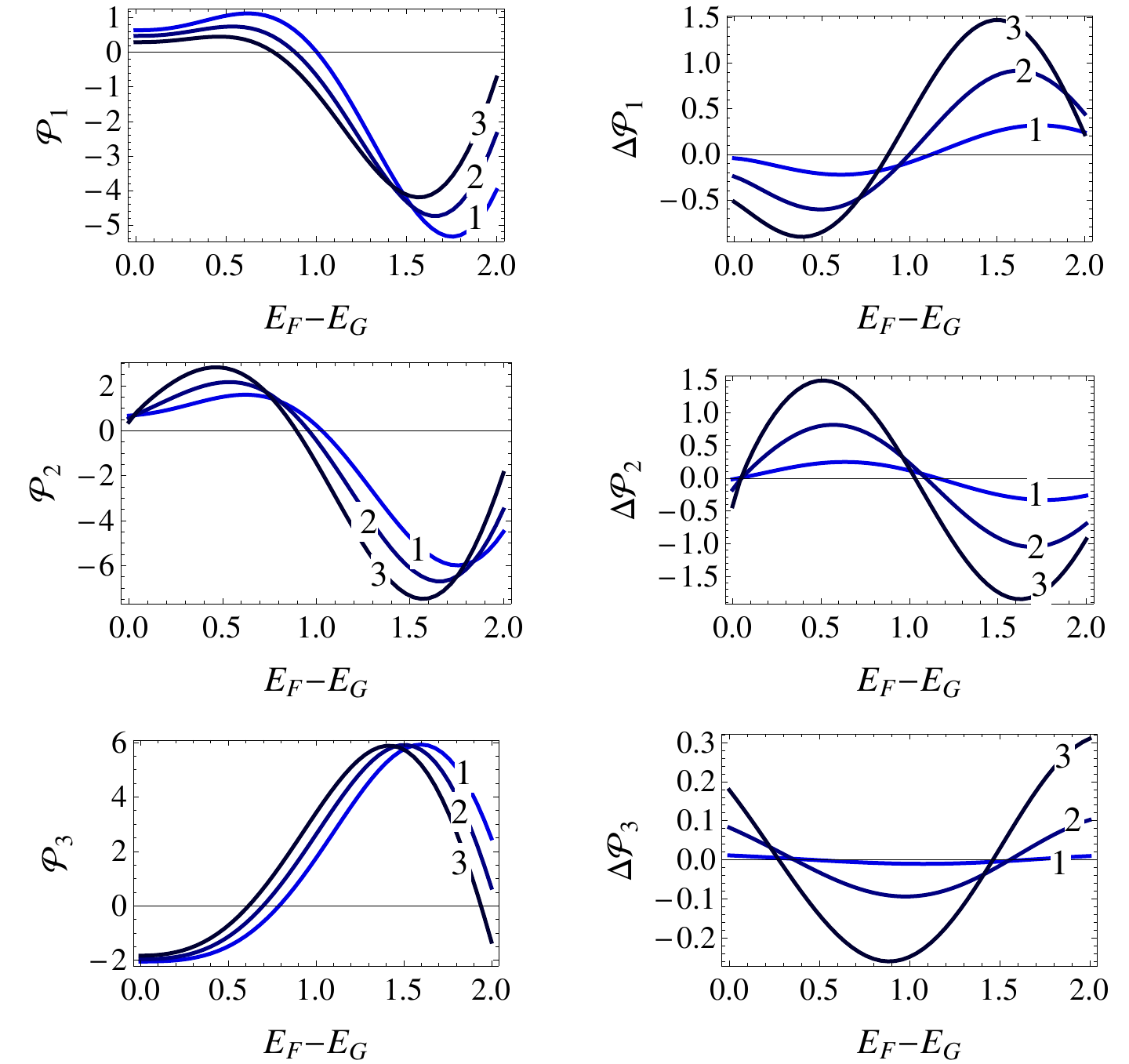}
\caption{\label{FIG:4} (Color online) Exchange interaction energies ${\cal P}_1$ (top panel), ${\cal P}_2$ (middle panel) and ${\cal P}_3$ (bottom panel) on the left column,
as well as the changes of them $\Delta{\cal P}_1$ (top panel), $\Delta{\cal P}_2$ (middle panel) and $\Delta{\cal P}_3$ (bottom panel) on the right column,
in doped graphene as functions of the Fermi energy $E_F-E_g$
for fixed $R_{AA}=R_{BB}=R_{AB} =1.0$. Curves $1,\,2,\,3$ correspond to the gap energy parameter $E_g = 0.1,\,0.3,\,0.5$ respectively.
Here, we measure  energy in  units of $\gamma_0$ and
distance in units of $R_{\mu\nu}=|\textbf{r}_\mu-\textbf{r}^\prime_\nu|/a$ for $\mu,\,\nu=A,\,B$.}
\end{figure}

Figure\ \ref{FIG:4} presents the calculated results for
${\cal P}_j(E_g, R_{\mu\nu})$ (left column) and
$\Delta {\cal P}_j(E_g, R_{\mu\nu})$ (right column) as
functions of $E_F-E_g$ with $R_{AA}=R_{BB}=R_{AB}=1$ and
various values of $E_g$ in doped graphene. From the left
column of this figure, as predicted before\,\cite{RKKYG5}
we find that the band-filling introduces an oscillation
 between  positive (ferromagnetic) and negative (antiferromagnetic)
 signs in ${\cal P}_1$, ${\cal P}_2$ and ${\cal P}_3$.
In addition, the zeros (non-magnetic) in ${\cal P}_1$,
${\cal P}_2$ and ${\cal P}_3$ are shifted leftward by increasing
the values of $E_g$. From $\Delta {\cal P}_1$, $\Delta {\cal P}_2$
and $\Delta {\cal P}_3$ displayed in the right column of
this figure, we further discover that the magnitude of
such oscillations is enhanced significantly in the presence
 of an energy gap for a graphene layer.
\medskip

\begin{figure}[ht]
\centering
\includegraphics[width=\textwidth]{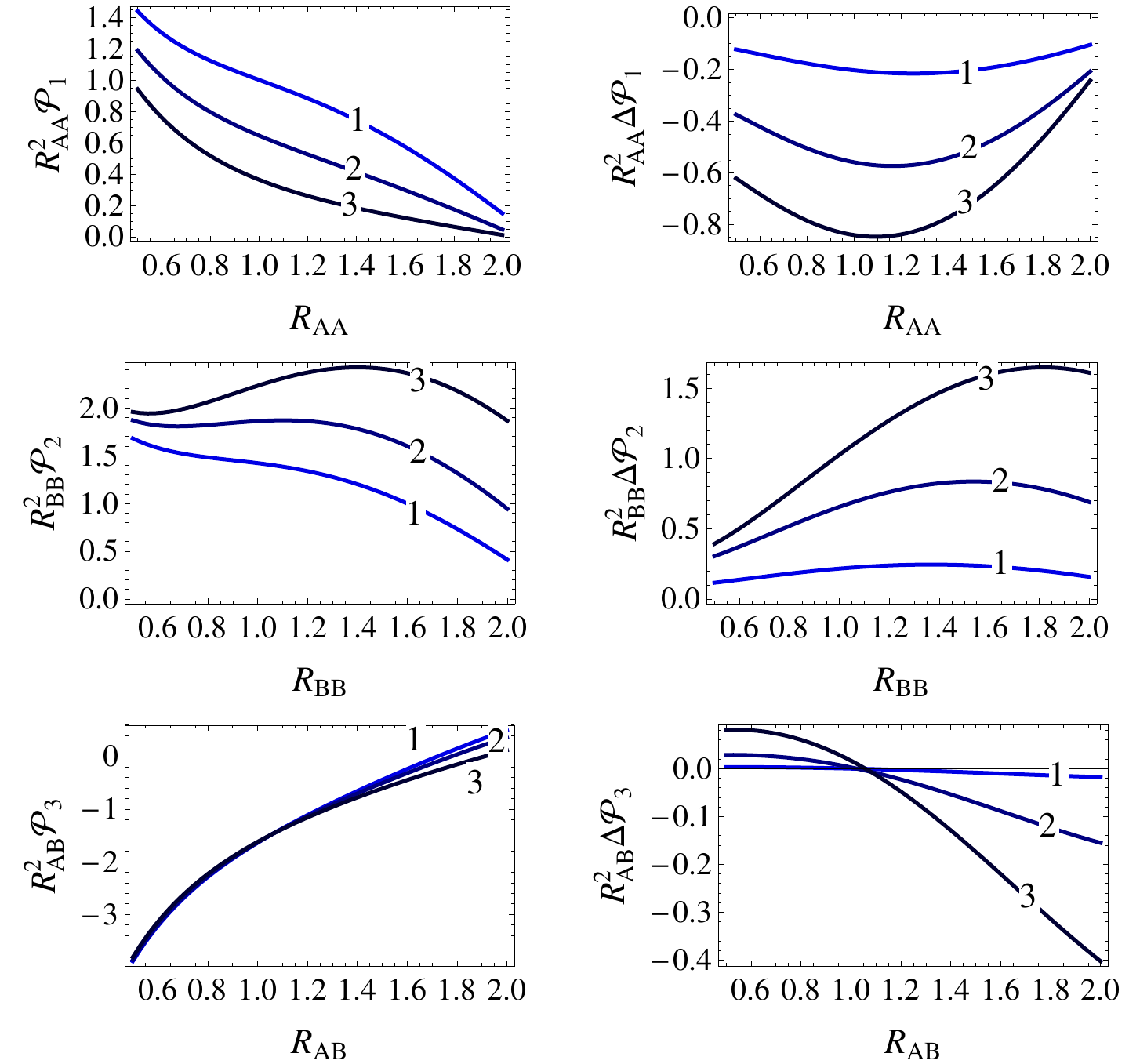}
\caption{\label{FIG:5} (Color online) Exchange interaction energies $R^2_{AA}{\cal P}_1$ (top), $R^2_{BB}{\cal P}_2$ (middle) and $R^2_{AB}{\cal P}_3$ (bottom) on the left column,
as well as the changes of them $R^2_{AA}\Delta{\cal P}_1$ (top), $R^2_{BB}\Delta{\cal P}_2$ (middle) and $R^2_{AB}\Delta{\cal P}_3$ (bottom) on the right column,
in doped graphene at $E_F-E_g=1$ as functions of the distance between two magnetic impurities. Curves $1,\,2,\,3$ correspond to  gap energy parameter $E_g = 0.1,\ 0.3,\ 0.5$, respectively.
Here, we measure  energy in  units of $\gamma_0$ and
distance in units of $R_{\mu\nu}=|\textbf{r}_\mu-\textbf{r}^\prime_\nu|/a$ for $\mu,\,\nu=A,\,B$.}
\end{figure}

We exhibit in Fig.\,\ref{FIG:5} our numerical results for
$R_{\mu\nu}^2{\cal P}_j(E_g, R_{\mu\nu})$ (left column) and
$R_{\mu\nu}^2\Delta {\cal P}_j(E_g, R_{\mu\nu})$ (right column)
as functions of the distance $R_{\mu\nu}$ between two magnetic
impurities with $E_F-E_g=1$ and various values of $E_g$ in
doped graphene. From the left column of this figure, we find
that both positive intra-SI exchange interactions
${\cal P}_1$  and ${\cal P}_2$ decrease with $R_{\mu\mu}$ but
the increase of $E_g$ suppresses  (enhances) ${\cal P}_1$ (${\cal P}_2$),
respectively. The mixed $1/R^2_{\mu\mu}$ and $1/R^3_{\mu\mu}$
dependence in ${\cal P}_{1,2}(E_g, R_{\mu\nu})$, as predicted in Ref.\,[\onlinecite{RKKYG5}] for doped graphene, is modified
dramatically. This clearly demonstrates the gap modulation
of the doping effect on the intra-SI indirect exchange interaction
between magnetic impurities in graphene. Moreover, ${\cal P}_3$
is found to increase with $R_{AB}$ with a negligible effect
from $E_g$, which is accompanied by a switching from
antiferromagnetic (negative) to ferromagnetic (positive)
beyond a threshold distance between a pair of magnetic impurities
in  the inter-SI exchange interaction. From the right-hand column
of this figure, we find that these exists a negative minimum
(positive maximum) for the gap-induced change
$\Delta {\cal P}_1$ ($\Delta {\cal P}_2$) as a function of
$R_{\mu\mu}$, respectively. Interestingly, the gap-induced change
$\Delta {\cal P}_3$ for the inter-SI exchange interaction shows up
a sign change with increasing $R_{AB}$ for all chosen finite values of $E_g$.

\section{Concluding Remarks}
\label{SEC:4}

\begin{figure}[ht]
\centering
\includegraphics[width=\textwidth]{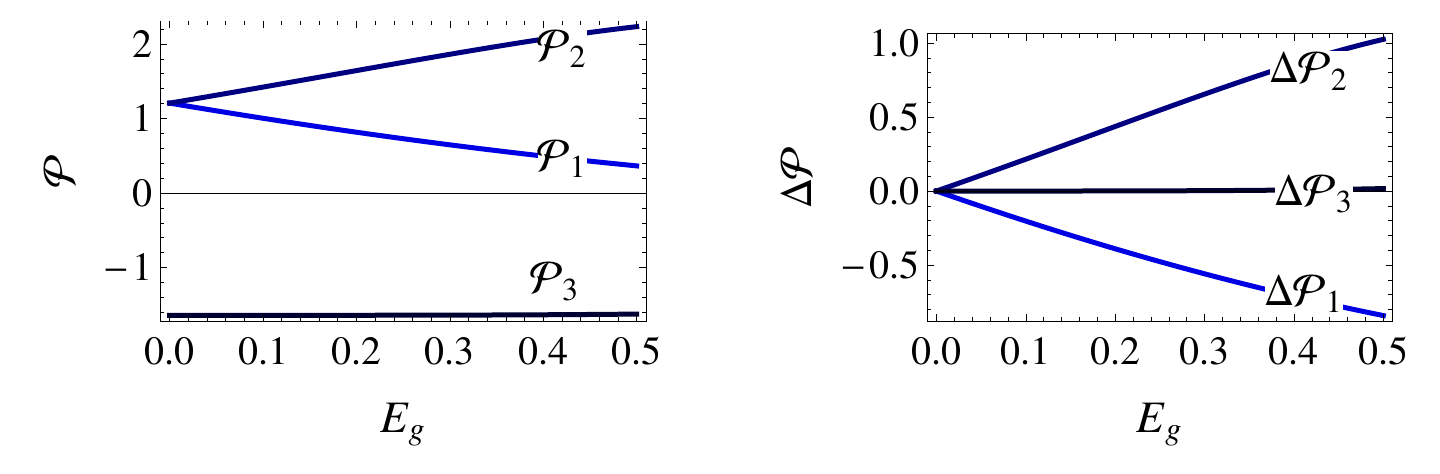}
\caption{\label{FIG:3} (Color online) Exchange interaction energies ${\cal P}_1$, ${\cal P}_2$ and ${\cal P}_3$ (left panel),
as well as the changes of them $\Delta{\cal P}_1$, $\Delta{\cal P}_2$ and $\Delta{\cal P}_3$ (right panel),
in doped graphene as functions of the half-gap $E_g$
for fixed $R_{AA}=R_{BB}=R_{AB} =1$ and $E_F-E_g=1$.
Here, we measure  energy in  units of $\gamma_0$ and
distance in units of $R_{\mu\nu}=|\textbf{r}_\mu-\textbf{r}^\prime_\nu|/a$ for $\mu,\,\nu=A,\,B$.}
\end{figure}

The RKKY interaction between a pair of magnetic impurities
located on lattice sites of monolayer half-filled gapped
graphene is theoretically investigated based on the lattice
Green's function formalism. In contrast with the case for
 gapless monolayer graphene, the RKKY interaction in gapped
 graphene possesses distinctive properties. Our numerical results
 in Fig.\,\ref{FIG:3} showed that although the presence of a gap
 does not change the ferromagnetic nature of the interaction
 within a chosen sublattice, it introduces an asymmetry between
 these two sublattices. We find that the larger the band gap,
 the smaller is the ferromagnetic energy for one of the two
 sublattices and the opposite behavior for another sublattice
 at the same time. The antiferromagnetic interaction is also
 reduced as the gap is increased for undoped graphene but not
 for doped graphene. Moreover, we found the gap modulation to the
 distance dependence of the RKKY interaction for both undoped ($1/R^3$)
 and doped (mixed $1/R^2$ and $1/R^3$) graphene, where $R$ is
 the distance between two magnetic impurities on sublattice sites.
Additionally, we showed that the doping-induced oscillations in
the magnetization within a sublattice or between sublattices can
be substantially modified by an energy gap in both magnitude and phase.
Some time ago, Gumbs and Glasser\,\cite{Gumbs-Glasser} showed
that the RKKY interaction in a metal
is modified by the presence of a surface and its long-distance
behavior is not given by a simple power law as in the bulk.
The results in the present paper again confirm that the RKKY
interaction may be significantly influenced by chemical
composition or the local environment of the material.

\acknowledgments
This research was supported by the contract \# FA9453-11-01-0263 of AFRL. DH would like
to thank the Air Force Office of Scientific Research (AFOSR) for its support.

\newpage

\end{document}